\documentclass[aps,pre,amsmath,amssymb,floatfix,superscriptaddress]{revtex4}
\usepackage{graphicx}
\usepackage[lofdepth,lotdepth]{subfig}
\usepackage{float}
\usepackage{epsfig}
\usepackage{inputenc}
\usepackage{bm}
\usepackage{gensymb}
\usepackage{mathtools}
\usepackage{array,multirow}
\usepackage{color}
\usepackage{hyperref}
\usepackage{url}
\usepackage{epstopdf}
\newcommand{\vect}{\mathbf}

\begin{document}

\title{Quantum structure of glasses and the boson peak:\\
a theory of vibrations}

\author{T.~R.~Cardoso}
\email{tatiana.cardoso@dfi.ufla.br, cardoso@ift.unesp.br}
\affiliation{Departamento de F\'isica, Universidade Federal de Lavras, \\
Caixa Postal 3037, 37200-000, Lavras-MG, Brazil\\}
\author{A.~Tureanu}
  \email{anca.tureanu@helsinki.fi}
\affiliation{ Department of Physics, University of Helsinki,\\
P.O. Box 64, 00014 Helsinki, Finland}

\begin{abstract}
We present a novel analytical model for glasses, starting from the first principle that the disorder in a glass mimics the disorder in a fluid. The origin of the boson peak is attributed to the intrinsically noncommutative geometry of the fluid disorder, which induces a van Hove singularity in the vibrational density of states. The universality of the model is exhibited by applying it to amorphous silicon, vitreous GeO$_2$ and Ba$_{8}$Ga$_{16}$Sn$_{30}$ clathrate, which show a remarkable agreement between the theoretical predictions for specific heat and the experimental data.
\end{abstract}

\maketitle
%\date{}

\section{Introduction}

Glasses are a peculiar state of matter whose mechanical rigidity is similar to that of crystalline solids, while the molecular disorder makes them akin to liquids \cite{berthier}. The significant differences in the thermodynamical and transport properties of glasses as compared to crystalline solids are generally attributed to a universal anomaly in the frequency spectra of atomic vibrations known as the “boson peak”. This appears at low energies as an excess in the vibrational density of states (VDOS) compared to the one predicted by the standard Debye theory, and manifests itself in measurements of specific heat and thermal conductivity, as well as scattering of electromagnetic radiation and neutrons.
The lack of a general theory for disordered solids based on first principles is a standing challenge in condensed matter physics.  The theoretical description of the anomalous behavior of disordered solids is a reccurent theme in the literature \cite{livro,berthier}. Interpretations of experimental studies seem to support various hypotheses:  the glass anomaly could be assigned to structural motifs with (quasi-)localized transverse vibrational
modes that resonantly couple with transverse phonons and thus
govern their dissipation \cite{shintani}, or to a broadening of the transverse acoustic van Hove singularity of the corresponding crystal \cite{chumakov}, and still
%This implies that the density of states does not contain extra modes in the low energy region. Consequently, the boson peak is not due to an excess of states, but rather to the piling up of the same number of states at lower energies than the TA peak, near the boundary of the pseudo-Brillouin zone.
to the presence of additional vibrational modes
which can be induced solely by structural disorder  \cite{Brink}.  There are mainly two classes in which the theoretical models can be grouped: i) models attributing the boson peak to quasilocalized modes, of non-acoustic nature \cite{Karpov,Laird} and ii) models with broadening of the crystalline van Hove singularity, like those based on randomly fluctuating elastic constants inserted in an otherwise crystalline arrrangement \cite{27}-\cite{29}. There is an impressive array of phenomenological models of glass structure and dynamics (for an inevitably incomplete selection, see Refs. \cite{21}-\cite{33}).
%
%Several examples of glass models include  localization
%of vibrations in phonon scattering caused by intrinsic density fluctuation domains \cite{21}; frequency resonance or localization of %phonons due to cohesive
%clusters\cite{25}; soft anharmonic potentials \cite{22}--\cite{24}; resonant or localized modes associated with locally favoured
%structures \cite{34}; a crossover
%between a minima-dominated phase (with phonons) and a
%saddle-point-dominated phase (without phonons) in the complex
%energy landscape \cite{30}; liquid-glass transitions in the
% the mode-coupling theory \cite{26,26'};  strongly anharmonic transitions between the
%local minima of the energy landscape\cite{31}; jammed packings of particles interacting with finite-range, repulsive potentials at %zero temperature \cite{33}.
%
However, none of these advances has allowed the derivation of a complete, widely-accepted theory of amorphous solids. Thus, the physical origin of two most outstanding anomalous behaviors displayed by glasses at intermediate temperatures ($1-40$ K), i.e., the excess of heat capacity and the boson peak, is still under debate.

In this work, we present a novel theory for the vibrations in amorphous solids, relying on the fundamental principle that the disorder in a glass is a reminiscence of the disorder in a fluid. The analytical model for glasses based on this hypohesis naturally contains the boson peak as a manifestation of an extended van Hove singularity.

\section{Fluids in Lagrangian description}

A fluid can be described in the Lagrangian picture (see, e.g., \cite{Mechanics}), in which one follows the motion of each individual fluid particle. The Lagrange coordinates are co-moving with the fluid.
%In the discrete description, the fluid dynamics is invariant under
%the renaming of the arbitrary particle labels.
Assuming that the fluid is a continuum, it has been shown \cite{Susskind-Bahcall, Susskind, Jackiw1} that the fluid dynamics encoded in the Lagrangian of the system is invariant under a reparametrization of the coordinates. In the discrete description, this is equivalent to invariance under renaming the arbitrary particle labels. Such
transformations are volume-preserving diffeomorphisms:
\begin{equation}\label{6}
\vect x\to \vect x'=\vect x+\delta\vect x, \ \ \ \  \delta \vect x=\vect f(\vect x),
\end{equation}
with
$
\nabla\cdot \vect f(\vect x)=0.
$
For simplicity, we consider a two-dimensional space. Then the latter condition can be written in terms of a scalar function $f$ as
\begin{equation}\label{8}
f_i(\vect x)=\epsilon_{ij}\frac{\partial f}{\partial x_j}.
\end{equation}  It has been proven (see also \cite{Jackiw2, Polyn}) that in this approach, the reparametrization symmetry can be re-cast in the form of noncommuting coordinates, namely by introducing additional Poisson brackets
\begin{equation}
\{x_{i},x_{j}\}=\theta _{ij},  \label{9}
\end{equation}%
with an arbitrary set of constants $\theta_{ij} $. The isotropy of the fluid is ensured through the arbitrariness of the elements of the matrix $\theta_{ij} $. Rescaling $f$ by $\theta^{-1}$, we can re-write $\partial x_i$ as
\begin{equation}\label{10}
\delta x_i=\theta_{ij}\partial_j f=\{x_i,f\}.
\end{equation}
The Poisson brackets \eqref{9} are invariant under the transformations \eqref{10}, the elements $\theta_{ij}$ remaining invariant under the reparametrization of coordinates
%. The proof is based on the fact that the Lie derivative of the tensor $\theta_{ij}$ with respect to the vector field $\vect f$ vanishes
\cite{Jackiw1}. The generalization to three- and higher-dimensional spaces is straightforward.

%The real ($\mathbf{X}$-space) configuration of the fluid with the density $
%\rho _{0}$ is considered as a equilibrium configuration, for which the
%potential has a minimum.

In the quantum treatment of the fluid, the Poisson bracket of coordinates
\eqref{9} becomes the commutator
\begin{equation}
\lbrack \widehat x_{i},\widehat x_{j}]=i\theta _{ij},  \ \ \ i,j=x,y,z,
\label{12}
\end{equation}%
(with an antisymmetric matrix $\theta_{ij}$), which has to be added to the usual canonical commutation relations
\begin{equation}
\lbrack \widehat x_{i},\widehat p_{j}]=i\hbar \delta _{ij},\ \ \ \ [\widehat p_{i},\widehat p_{j}]=0.  \label{13}
\end{equation}
The theory based on the commutators \eqref{12} and \eqref{13} is customarily called noncommutative quantum mechanics.

To the matrix $\theta_{ij}$ we can associate a vector $\vec \theta$, whose components are given by $\theta_i=\epsilon_{ijk}\theta_{jk}$. We note that there are volume-preserving diffeomorphisms which leave invariant
simultaneously $\vec \theta $ and the surface density of particles $\sigma _{0}$ in any plane of the fluid. We expect a relation between these two invariants, which can be derived by analogy with
usual quantum mechanics. In the latter case, the commutation relation
$[\widehat x,\widehat p]=i\hbar $ leads to a quantization of the
two-dimensional phase space in cells of area $2\pi \hbar $. In the case of noncommuting
coordinates, the commutator \eqref{12} leads to a quantization of the
reference configuration space in cells of area $2\pi \theta $. We attribute to this basic
quantum of area the meaning of surface ``occupied'' by a single particle \cite{Susskind}. But  the inverse density of particles has the same
significance, leading to the
relation
\begin{equation}
\theta =\frac{1}{2\pi \sigma _{0}}.  \label{14}
\end{equation}
Incidentally, this approach has been applied to the quantum Hall fluid, providing an alternative description to Laughlin's theory by a noncommutative Chern--Simons quantum field theory \cite{Susskind}.

Let us dwell for a while on the physical significance of the commutator \eqref{12}: its presence in the quantum algebra introduces an additional {\it uncertainty relation}, which manifests itself as a ``blurriness'' of space points. Precise localization, even theoretical, of the particles becomes impossible.
%Within an area $\sigma_0$ in any plane of the fluid, a representative particle can be anywhere.
This suggests an intrinsic  ``disorder''. As we shall see further, the dynamics is profoundly affected, by the coupling of vibrational modes specific to noncommutative quantum mechanics.

The dynamics in noncommutative quantum mechanics is described by the Schr\"odinger equation for $N$ degrees of freedom, with the Hamiltonian
\begin{equation}\label{Hamilt_gen}
\widehat H \equiv \sum_i^N \frac{\widehat p_i \widehat p_i}{2M}+ V\left( \widehat x_1,\ldots, \widehat x_N\right),
\end{equation}
where the canonical coordinates $\widehat x_i,\widehat p_i, i=1,\ldots,N$ satisfy the extended Heisenberg algebra \eqref{12}-\eqref{13}. The mathematical manipulations are considerably simplified by noting (see, e.g., \cite{CDP_98, Bigatti_Susskind}) that the shifted coordinates
\begin{eqnarray}\label{Bopp}
\widehat X_{i} = \widehat x_i + \frac{1}{2\hbar} \theta_{ij}\widehat p_{j},\ \ \ \
\widehat{P}_{i}= \widehat p_{i}
\end{eqnarray}
satisfy the usual Heisenberg algebra $[\widehat X_i,\widehat P_j]=i\hbar\delta_{ij}$, $[\widehat X_i,\widehat X_j]=0$, $[\widehat P_i,\widehat P_j]=0$ (summation over the repeated index is assumed in \eqref{Bopp}). This allows a physical interpretation of the $\theta$-term in \eqref{Bopp} as a quantum shift operator  \cite{Anca}, namely, a translation in space by $\frac{1}{2\hbar} \theta_{ij}\widehat p_{j}$, while $X_{i}$ is the classical geometrical coordinate. This technique has been used for deriving noncommutative space corrections to various quantum mechanical phenomena, for example, the Lamb shift \cite{Tureanu}. If the potential energy in \eqref{Hamilt_gen} is of the harmonic oscillator type, then, when the shift \eqref{Bopp} is applied to $H \left(\widehat{ x},\widehat{ p}\right)$, one obtains a Hamiltonian $H (\widehat X,\widehat P)$ consisting of a sum of usual quantum mechanical harmonic oscillators plus an additional interaction term proportional to the noncommutativity parameter.

%However, a three-dimensional antisymmetric matrix has determinant zero,
%which means that the coordinates can be rotated in such a way that only two
%of the new coordinates do not commute among themselves, i.e.
%\begin{equation}
%\theta _{ij}^{\prime }=%
%\begin{bmatrix}
%0 & 0 & 0 \\
%0 & 0 & \theta ^{\prime } \\
%0 & -\theta ^{\prime } & 0%
%\end{bmatrix}
%\label{19}
%\end{equation}%
%The plane of the noncommuting coordinates is orthogonal on the vector $%
%\mathbf{\theta }$, whose direction can be easily calculated from \eqref{18}
%to be along one diagonal of the cubic cell of the lattice. When one applies
%the formula \eqref{14}, it is this $\theta ^{\prime }$ of \eqref{19} which
%is relate to the ``surface'' density of particles in the noncommutative
%plane.

\section{Model for glassy materials: reduced specific heat and the boson peak}

The glass structure model we propose avails itself of the noncommutative fluid picture. Glasses are not normal liquids due to their {\it rigidity}, nor regular solids, due to their {\it disorder}.
%Therefore the model has to combine these two features.
The rigidity is introduced by means of a simple cubic lattice. The disorder reminiscent of the fluid which was quenched into the glass will be implemented through the noncommutative quantum algebra \eqref{12}-\eqref{13}.

We assume the glass composed of a simple cubic lattice of neutral atoms of mass $M$, with the unit cell vector $a$. The dynamics of the disordered lattice is described by a noncommutative harmonic oscillator potential function and we consider harmonic interactions only between the first neighbors of the atoms in a lattice. We take the vector $\vec\theta$ with equal projections denoted by $\theta$ on the coordinate axes, i.e. on the directions of the edges of the unit cells. Let us analyze the atom indexed by the integer labels ${lmn}$. The quantum coordinate operators which define it are $\widehat x_{lmn}={l}\,a+\widehat u^x_{lmn}$, $\widehat y_{lmn}={m}\,a+\widehat u^y_{lmn}$, $\widehat z_{lmn}={n}\,a+\widehat u^z_{lmn}$, where $\widehat u^i$, with $i= x,y,z$, denote generically the displacement operators from the lattice site in the corresponding direction. The Hamiltonian
governing the time evolution
of the atom $({lmn})$ is then
\begin{eqnarray} \label{ham}
\widehat H_{lmn} & =&\sum_{i=x,y,z}\frac{1}{2M}\left(\widehat{p}^i_{lmn}\right)^{2} \\
& +&\frac{M\omega_{0}^{2}}{2}\left[\left(\widehat u^x_{lmn}-\widehat u^x_{l-1mn}\right)^{2}+\left(\widehat u^x_{l+1mn}-\widehat u^x_{lmn}\right)^{2}\right]\cr
& +&\frac{M\omega_{0}^{2}}{2}\left[\left(\widehat u^y_{lmn}-\widehat u^y_{lm-1n}\right)^{2}+\left(\widehat u^y_{lm+1n}-\widehat u^y_{lmn}\right)^{2}\right]\cr
& +&\frac{M\omega_{0}^{2}}{2}\left[\left(\widehat u^z_{lmn}-\widehat u^z_{lmn-1}\right)^{2}+\left(\widehat u^z_{lmn+1}-\widehat u^z_{lmn}\right)^{2}\right]\nonumber,
\end{eqnarray}
where $\widehat{p}_{lmn}^{i}$ is the momentum canonically conjugated to the displacement $\widehat u^i_{lmn}$ and $\omega_0$ is the usual harmonic oscillator frequency.

In order to make sure that the disorder effects are not doubly-counted, nor washed out, we make an additional assumption about the dynamics of the atoms on the lattice. Namely, we consider an alternation of ordered and disordered atoms. By {\it ordered atoms}, we mean atoms whose quantum coordinates and momenta satisfy the usual Heisenberg algebra, i.e. whose coordinate operators commute. By {\it disordered atoms}, we mean atoms whose coordinates and momenta satisfy the noncommutative space algebra \eqref{12}-\eqref{13}. The arrangement of the lattice is such that one ordered atom is surrounded by disordered first neighbors and vice-versa. Due to the elastic couplings, the equations of motion of the so-called ordered atoms will be also influenced by the disorder of their neighbors, such that the lattice as a whole will be disordered.

Let us specify the Hamiltonian \eqref{ham} for the two types of atoms. Considering that the atom $lmn$ is a disordered one, which suffers itself the effects of the noncommutativity of coordinates, we perform the shift \eqref{Bopp} by replacing in \eqref{ham}
\begin{eqnarray}
\widehat{u}^i_{lmn}&=&\widehat U^i_{lmn}-\frac{1}{2\hbar}\theta_{ij}\widehat P_{lmn}^j,
\end{eqnarray}
while for all the displacements of the nearest neighbors we have $\widehat u^i_{l+1mn}= \widehat U^i_{l+1mn}$ etc.

If the generic atom $lmn$ is a ordered one, the shifts \eqref{Bopp} in the Hamiltonian \eqref{ham} have to be performed for the coordinates of the neighbors, i.e.
\begin{eqnarray}
\widehat{u}^x_{l\pm1mn}=\widehat U^x_{l\pm1mn}-\frac{1}{2\hbar}\theta_{xj}\widehat P_{l\pm1mn}^{j}, \ \  j=y,z,
\end{eqnarray}
while $\widehat u^i_{lmn}=\widehat U^i_{lmn}$, for $i=x,y,z$.
In both cases,
$
\widehat{p}_{lmn}^{i}=\widehat P_{lmn}^{i} .
$

The equations of motion are obtained by applying the Heisenberg equations
$
i\hbar\frac{d\widehat O\left(t\right)}{dt}=\left[\widehat O\left(t\right), H\right],
$
taking into account the canonical commutation relations
\begin{eqnarray}
\left[\widehat U_{lmn}^{i},\widehat P_{l^{\prime }m^{\prime }n^{\prime }}^{i'}\right]
=i\hbar\delta^{ii'}\delta_{ll^{\prime }}\delta_{mm^{\prime }}\delta_{nn^{\prime }},\cr
\left[\widehat U_{lmn}^{i},\widehat U_{l^{\prime }m^{\prime }n^{\prime }}^{i'}\right]=\left[\widehat P_{lmn}^{i},\widehat P_{l^{\prime }m^{\prime }n^{\prime }}^{i'}\right]=0.
\end{eqnarray}

The equations of motion (see Appendices) are differential equations which couple displacements in all three directions.   We introduce the notation \textcolor{black}{$\omega_{\theta}= \frac{\hbar}{M \theta}$} and $R=\omega_0/\omega_\theta$. The latter is a dimensionless expansion parameter, proportional to $\theta$. We have retained only terms up to the second order in $R$, as this parameter is expected to be very small. This expectation will be fully justified later, in the comparison with the experimental data.

Further, we use Born's approach to the vibrations of a lattice, expanding in Fourier series the operators
\begin{equation}\label{Fourier_solution}
\hat U^i_{lmn}=\sum_{\vect k}\hat{\cal U}^{i}_{\vect k}\exp \left[ i\left( \omega_{\vect k} t+a(lk_x+mk_y+n k_z) \right)\right],
\end{equation}
and applying the Born--von K\'{a}rm\'{a}n boundary conditions (see, e.g., \cite{Kantorovich}). In the above expression, $k_i$ represent the wave vector projections and $\omega_{\vect k}$ the mode vibration frequencies. We obtain the dispersion relations by inserting the expansion \eqref{Fourier_solution} into the equations of motion.

%At this point, we have to consider the isotropy issue.
So far, the model has two sources of anisotropy: the crystal lattice and the vector  $\vec \theta$. As far as the noncommutativity is concerned, the disorder effect is manifest only in the plane orthogonal to $\vec\theta$, and not in the direction of $\vec \theta$. In order to render the model isotropic, we choose as representative axis one of the coordinates axes (e.g. $Oz$), and determine the dispersion relations for it, $\omega_{k_z}(k_z)$. With our choice of $\vec\theta$ having equal projections on the coordinate axes, we insure that the disorder induced by noncommutativity is similar along $Oz$ and in the plane orthogonal to it. Subsequently, we replicate the $Oz$ axis by rotational symmetry to all the directions of the Cartesian frame, i.e. $\omega_{k_z}(k_z)\to \omega_{\vect k}(|\vect k|)$, by replacing in $\omega_{k_z}(k_z)$ everywhere $k_z$ by $|\vect k|$.

The isotropised model gives the following dispersion relations:
\begin{equation}
 k _{x}^{2}+k _{y}^{2}+k _{z}^{2} =\frac{1}{a^2}\frac{\omega_{\vect k}
^{2}\left({\omega_{\vect k} ^{2}}/{\omega _{0}^{2}} -3R^{2}\right) }{\omega_{\vect k} ^{2}\left( 1+R^{2}\right)-\omega
_{0}^{2}R^{2} } , \label{disp_D}
\end{equation}
for disordered atoms, and
\begin{equation}
 k _{x}^{2}+k _{y}^{2}+k _{z}^{2} =\frac{1}{a^2}\frac{\omega_{\vect k}
^{2}\left({\omega_{\vect k} ^{2}}/{\omega _{0}^{2}} -3R^{2}\right) }{\omega_{\vect k} ^{2}-\omega
_{0}^{2}R^{2} } , \label{disp_O}
\end{equation}
for ordered atoms. These expressions are obtained in the small-angle approximation (i.e. $\sin(ak_i)\approx ak_i$), but they are valid with excellent accuracy over the whole Brillouin zone, due to the smallness of the $R$-parameter (for $R=0.1$, the ratio between the approximate and the exact values of the frequency at the border of the Brillouin zone, where the discrepancy is maximal, is about 1.5 \%).	 We obtain three degenerated acoustic and three optical branches, the latter being a pure noncommutativity effect.

Due to the rotational symmetry, the VDOS is easily derived from \eqref{disp_D} and \eqref{disp_O} using standard methods, with the result:
\begin{equation}\label{g_glass}
g_{glass}\left( \omega \right) =g_{O}\left( \omega \right) \left[ 1+\frac{%
\omega ^{2}}{\omega _{0}^{2}R^{2}}\left\vert \frac{\frac{\omega ^{2}}{\omega
_{0}^{2}R^{2}}-2-\frac{3}{2}\frac{\omega ^{2}}{\omega _{0}^{2}}}{\left( 1-%
\frac{\omega ^{2}}{\omega _{0}^{2}R^{2}}\right) ^{2}}\right\vert \right],
\end{equation}%
where
\begin{equation}\label{g_ordered}
g_{O}\left( \omega \right) =3\frac{V}{(2\pi)^2}\frac{1}{a^3}\frac{\omega ^{2}}{\omega _{0}^{3}}\sqrt{%
\left\vert \frac{3-\frac{\omega ^{2}}{\omega _{0}^{2}R^{2}}}{1-\frac{\omega
^{2}}{\omega _{0}^{2}R^{2}}}\right\vert }
\end{equation}%
is the contribution to VDOS from the ordered atoms. We note the divergence in
VDOS, i.e. a van Hove singularity, which occurs for $\omega _{\mathrm{div}}=\omega _{0}R$. In the limit $\theta \rightarrow 0$, we recover the VDOS of a usual simple cubic lattice with one atom per cell.

The reduced specific heat is determined from the following expression (see, e.g., \cite{Kantorovich})
\begin{equation}\label{specific_heat}
\frac{C}{T^3}=\int_0^{\omega_{\rm{max}}} \frac{ \hbar^2 \omega^2}{k_{B}T^5}\frac{N_A}{Z} \frac{e^{\hbar\omega/k_{B}T}}{(e^{\hbar\omega/k_{B}T}-1)^2} g_{glass}(\omega) d\omega,
\end{equation}
in which $Z$ is the number of formula units per unit cell (in our model, $Z=1$) and $k_B$ is the Boltzmann constant. We take $\omega _{\mathrm{max}%
}=\sqrt 3\omega _{0}R$ as the frequency of the optical branches at $\vect k=0$.

\begin{figure}[h]
\subfloat[Subfigure 1 list of figures text][Amorphous silicon]  {\
\includegraphics[width=6cm]{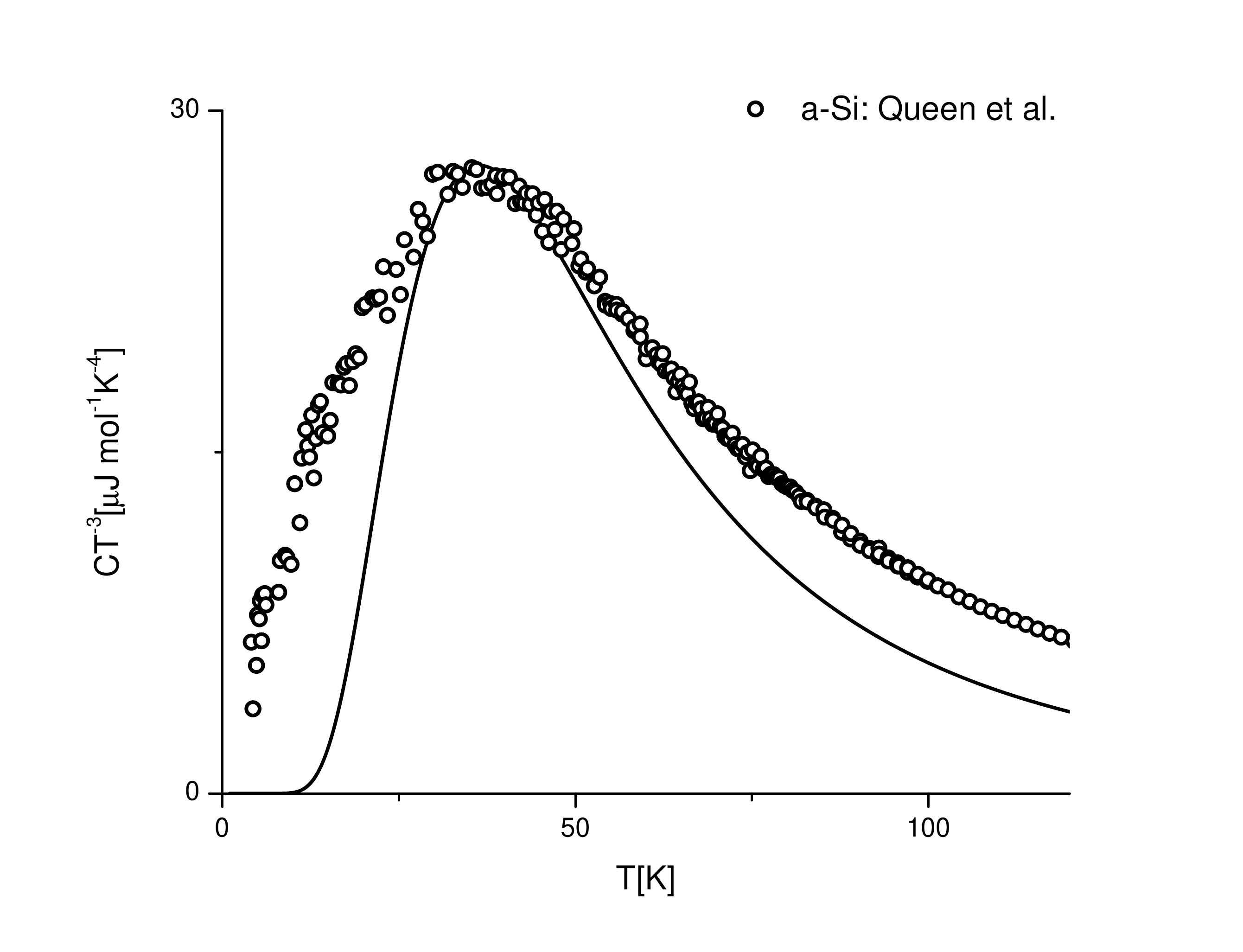}} ~
\subfloat[Subfigure 1 list of figures text][Amorphous germanium dioxide]  {\
\includegraphics[width=6cm]{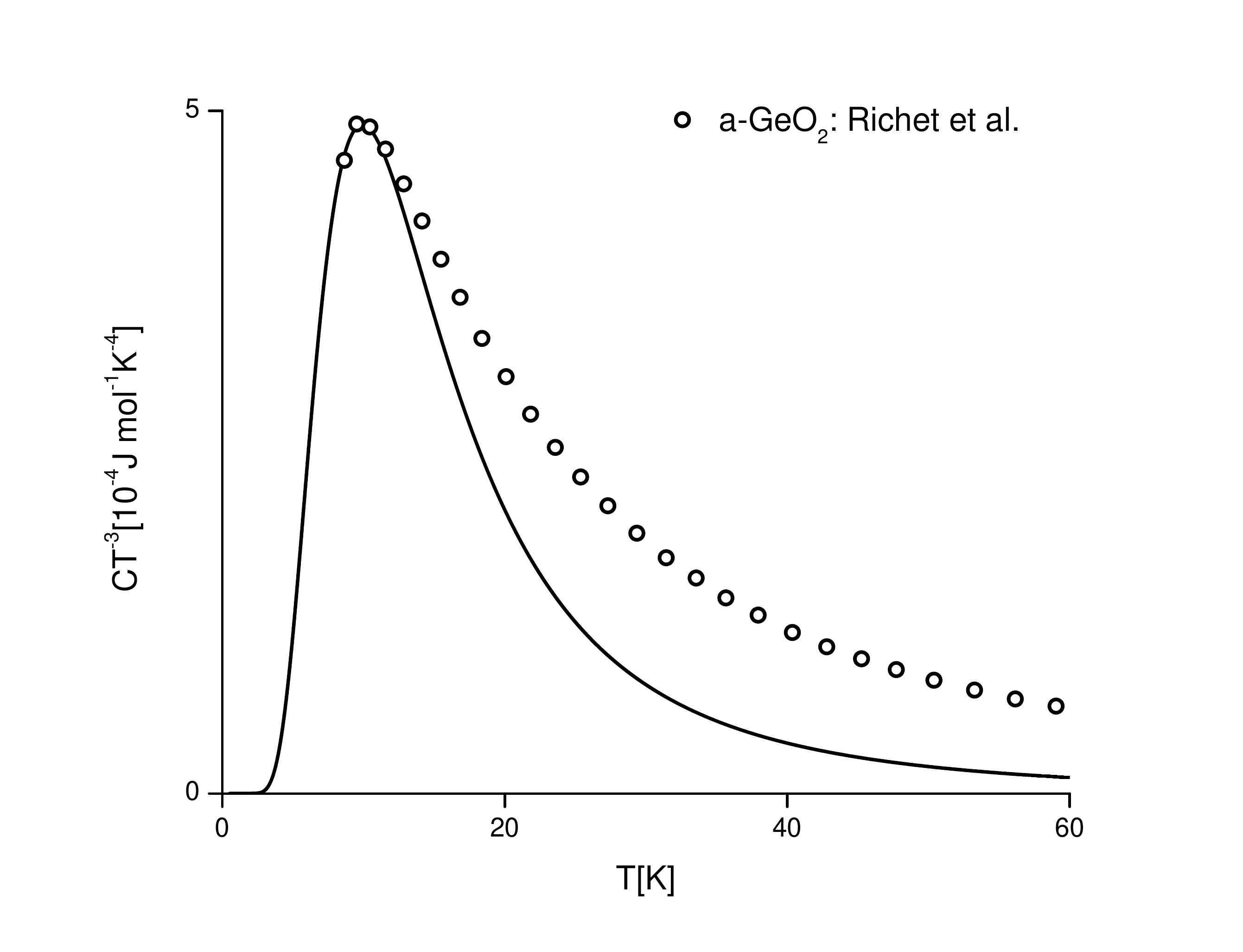}} ~
\subfloat[Subfigure 1 list of figures text][$\alpha -n-$Ba$_{8}$Ga$_{16}$Sn$_{30}$ clathrate]  {\
\includegraphics[width=6cm]{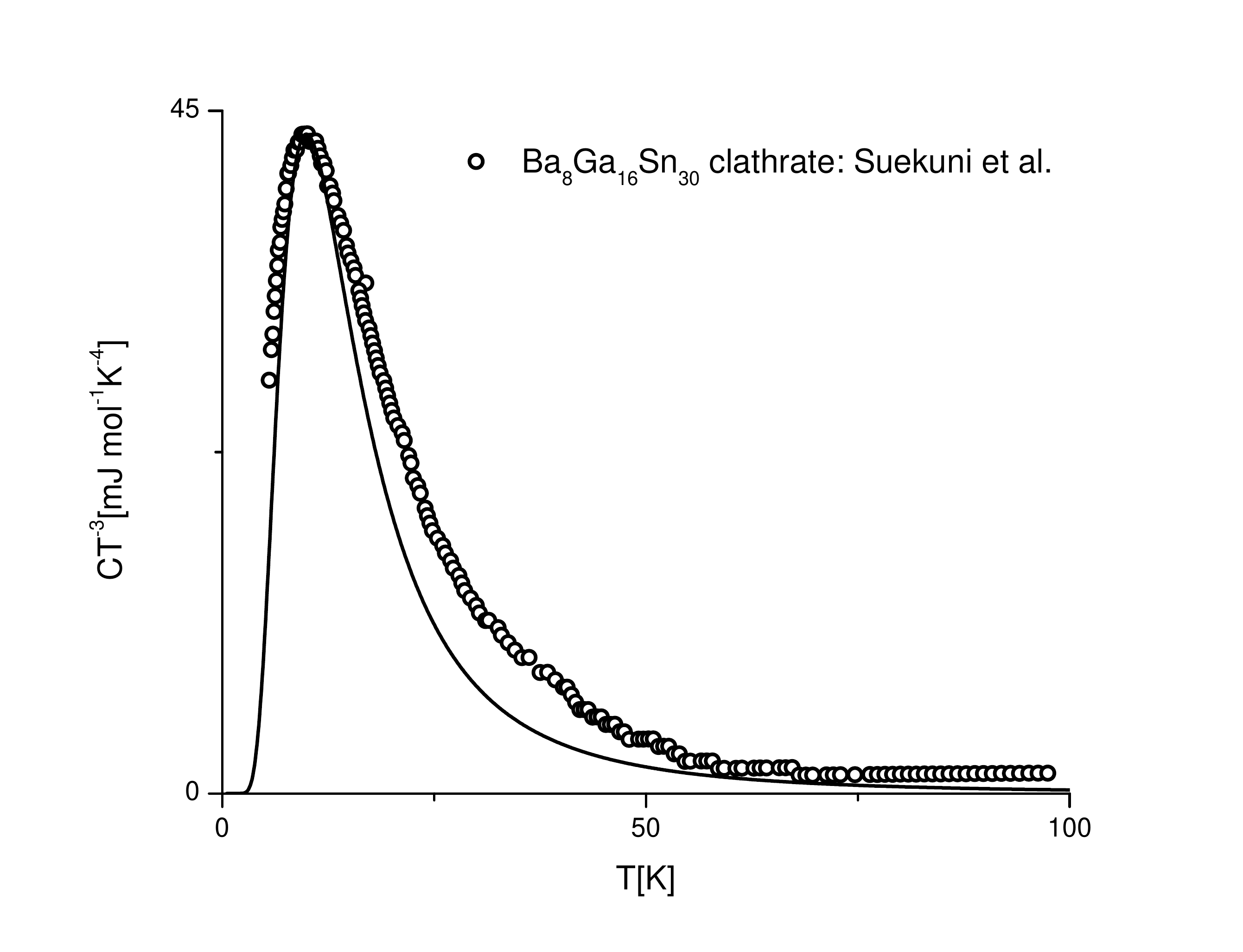}}
\caption{\textbf{Experimental data versus theoretical prediction.} Experimental data for (a) amorphous silicon ($a$-Si) \cite{queen}; (b) amorphous germanium dioxide ($a$-GeO$_{2}$) \cite{GeO2}; (c) $\alpha -n-$Ba$_{8}$Ga$_{16}$Sn$_{30}$ clathrate \cite{BGS}, respectively, given by the empty balls. The theoretical
prediction according to \eqref{specific_heat} is represented by the solid line.} \label{FigureA}
\end{figure}

The motif of the disordered lattice can be a single atom, a small molecule, a protein or any combination thereof. Hereafter, three distinct disordered materials, amorphous silicon ($a$-Si) \cite{queen}, vitreous GeO$_2$ \cite{GeO2} and $\alpha -n-$Ba$_{8}$Ga$_{16}$Sn$_{30}$ clathrate \cite{BGS}, are considered. The model has two free parameters: the characteristic frequency  $\omega_0$ and the noncommutativity parameter $\theta$.  They are determined by fitting the theoretical curve \eqref{specific_heat} to the experimental data, such that the frequency and reduced specific heat at the peak match (see table \ref{TableA}). In Fig.~\ref{FigureA} is shown the remarkable agreement around the peak between the experimental curves and the theoretical predictions of our model based on liquid-type disorder effects.

Let us recall the physical meaning of the parameter $\theta$: according to \eqref{14}, $2\pi\theta=\frac{1}{\sigma_0}$ is the area ``occupied'' by a particle in the quantized configuration space, or the area within which the uncertainty in the position of the particle is significant. On the other hand, the mass density of the glass leads to a value of the average interatomic distance $a$, which we consider to be the lattice spacing for our model. We note that the ratio $2\pi\theta/a^2$ is of the same order of magnitude for the analyzed glasses (see table \ref{TableA}). It would be interesting to investigate whether this regularity is valid for other glasses as well and find its physical significance.

\begin{table}
\begin{tabular}{|c|c|c|c|c|c|c|c|c|}
\hline
Glassy material & $R$ & $\omega _{0}\left( \times 10^{14}\right) $ & $\omega
_{\theta }\left( \times 10^{15}\right) $ & $\omega _{\mathrm{div}}\left(
\times 10^{12}\right) $ & $2\pi \theta \left( \times 10^{-24}\right) $ & $%
N\left( \times 10^{28}\right) $ & $a^{2}\left( \times 10^{-20}\right) $ & $%
T_{\mathrm{peak}}$ \\ \hline
$a$-Si & $0.0478$ & $4.916$ & $10.285$ & $23.500$ & $1.382$ & $4.710$ & $%
5.523$ & $35$ \\ \hline
$a$-GeO$_{2}$ & $0.0668$ & $1.003$ & $1.501$ & $6.700$ & $2.518$ & $2.655$ &
$8.066$ & $10$ \\ \hline
$\alpha n$BGS clathrate & $0.290$ & $0.241$ & $0.083$ & $7.000$ & $31.504$ &
$0.392$ & $34.106$ & $10.32$ \\ \hline
\end{tabular}
%\begin{flushleft}
\caption{Characteristic parameters for various glasses. All frequencies in rad s$^{-1}$. The value of $\theta $ (in m$^{-2}$) is determined from the relation $\omega _{\mathrm{div}%
}=\omega _{0}R$, using the experimental curves. $N$ is the number density in m$^{-3}$, whereas $a$ is the lattice spacing for our model in m$^{2}$. $T_{\mathrm{peak}}$ is the temperature at the peak of the reduced specific heat in K.}
%\end{flushleft}
\label{TableA}
\end{table}

\section{Conclusions and outlook}

In summary, an amorphous solid can be interpreted as a system with a frozen-in liquid-type of disorder, implemented mathematically as a noncommutative algebra of coordinate operators. Intuitively, this is equivalent to a blurriness or uncertainty in the positions of the particles that form the glass. However, this is not a simple positional disorder, as the delocalization depends essentially on the momenta of the particles -- a feature specific to noncommuting coordinates. The quantum mechanical model we propose is developed from first principles and permits the derivation of {\it analytic} formulas for the density of states and specific heat of the glass. Other important features of the model are its simplicity and the very small number of free parameters ($\omega_0$ and $\theta$, the former connected to the electromagnetic interactions among the glass particles and the latter being a measure of the disorder). This new theory naturally accounts for the excess of heat capacity and the boson peak phenomena (see Fig. \ref{FigureA}), which are manifestations of a pronounced divergence in the density of states in the acoustic branches, i.e. a van Hove singularity. The universality of the model is  confirmed by the excellent agreement between the theoretical predictions and the experimental specific heat data around the peak for an array of diverse glasses.

Glasses are complex systems, and their quantum behaviour at different temperatures is dominated by different aspects of their structure and dynamics. We remark that the departure of the experimental curves from the analytic curve is natural at temperatures further away from the peak, since formula for VDOS \eqref{g_glass} includes only the contribution of acoustic modes of the noncommutative simple cubic lattice with nearest-neighbor interactions. The model can be refined by including the next-to-nearest neighbor couplings, in which case we expect to see the peak in the transverse acoustic branch. It is also interesting to study whether adopting in the model the actual lattice type of the crystal counterpart of the analyzed glasses would improve the agreement with experimental data. A more
detailed analysis of the structural and dynamical aspects of the model is in progress  \cite{work in progress}.
%Further confirmations have to include the comparison of theoretical predictions with data for other quantities whose the anomalous behaviour is assigned to the boson peak, like, for example, the thermal conductivity.

\section*{Acknowledgements}
The authors are particularly grateful to R. Bufalo and M. Chaichian for
valuable discussions and advice.
We are also
grateful to F. Hellman for kindly sharing the
experimental data for amorphous silicon.
T.~R.~Cardoso thanks  B.~M.~Pimentel for discussions on noncommutative field theories and
H.~S.~Martinho for discussions on the structure of glasses and, in particular, for drawing her attention to the clathrate literature and data. The financial support of FAPESP through Project no. $2016/03921-8$, and the Academy of Finland through
the Projects no. 136539 and 272919 is acknowledged.

\appendix

\section{Noncommutative normal modes of vibration}

Assuming harmonic interactions among the nearest neighbors of a neutral
monatomic simple cubic lattice, the dynamics in noncommutative quantum
mechanics for normal modes of vibration of a simple cubic lattice is
described by the Schr\"{o}dinger equation for $N$ degrees of freedom. Our
starting point is the Hamiltonian governing the time evolution of the atom $%
(lmn)$:
\begin{eqnarray}  \label{ham}
\widehat H_{lmn} & =&\sum_{i=x,y,z}\frac{1}{2M}\left(\widehat{p}%
^i_{lmn}\right)^{2} \\
& +&\frac{M\omega_{0}^{2}}{2}\left[\left(\widehat u^x_{lmn}-\widehat
u^x_{l-1mn}\right)^{2}+\left(\widehat u^x_{l+1mn}-\widehat
u^x_{lmn}\right)^{2}\right]\cr & +&\frac{M\omega_{0}^{2}}{2}\left[%
\left(\widehat u^y_{lmn}-\widehat u^y_{lm-1n}\right)^{2}+\left(\widehat
u^y_{lm+1n}-\widehat u^y_{lmn}\right)^{2}\right]\cr & +&\frac{M\omega_{0}^{2}%
}{2}\left[\left(\widehat u^z_{lmn}-\widehat
u^z_{lmn-1}\right)^{2}+\left(\widehat u^z_{lmn+1}-\widehat
u^z_{lmn}\right)^{2}\right] ,  \notag
\end{eqnarray}
where $\widehat{p}_{lmn}^{i}$ is the momentum canonically conjugated to the
displacement $\widehat u^i_{lmn}$ and $\omega_0$ is the usual harmonic
oscillator frequency. The quantum algebra satisfied by the canonical
variables is:
\begin{eqnarray}
\left[\widehat u_{lmn}^{i},\widehat p_{l^{\prime }m^{\prime }n^{\prime
}}^{i^{\prime }}\right] &=&i\hbar\delta^{ii^{\prime }}\delta_{ll^{\prime
}}\delta_{mm^{\prime }}\delta_{nn^{\prime }},\cr \left[\widehat
u_{lmn}^{i},\widehat u_{l^{\prime }m^{\prime }n^{\prime }}^{i^{\prime }}%
\right]&=&i\theta_{ii^{\prime }},\cr \left[\widehat p_{lmn}^{i},\widehat
p_{l^{\prime }m^{\prime }n^{\prime }}^{i^{\prime }}\right]&=&0.
\end{eqnarray}

As explained earlier, the model consists of an alternation of ordered
and disordered atoms in a simple cubic lattice. \textit{Ordered atoms} mean
atoms whose quantum coordinates and momenta satisfy the usual Heisenberg
algebra, i.e., whose coordinates commute, while the \textit{disordered atoms}
have coordinates and momenta satisfying the noncommutative space algebra.
(There are ``two species'' of atoms in this case, like, for example, in a
NaCl lattice, just that the atoms are identical, but half are ordered and
half are not.)

We should emphasize that the distinction \textit{ordered/disordered atoms}
is a matter of semantics, because \textit{all} atoms suffer the effects of
the noncommutativity of coordinates either directly or through the dynamical
couplings. The lattice, as a whole, will be disordered.

\section{Equations of motion and dispersion relations for disordered atoms%
}

Considering that the atom $lmn$ is a disordered one, which suffers itself
the effects of the noncommutativity of coordinates, we perform in \eqref{ham}
the shift
\begin{eqnarray}
\widehat{u}^i_{lmn}&=&\widehat U^i_{lmn}-\frac{1}{2\hbar}\theta_{ij}\widehat
P_{lmn}^j,
\end{eqnarray}
while for all the displacements of the nearest (ordered) neighbors we have
\begin{eqnarray}
\widehat u^x_{l\pm1mn}= \widehat U^x_{l\pm1mn},\cr \widehat u^y_{lm\pm1n}=
\widehat U^y_{lm\pm1n},\cr \widehat u^z_{lmn\pm1}= \widehat U^y_{lmn\pm1}.
\end{eqnarray}
For the momenta, $\widehat{p}_{lmn}^{i}=\widehat{P}_{lmn}^{i}$, with $i=x,y,z
$. The shifted displacements obey the canonical commutation relations%
\begin{equation}
\left[ \widehat{U}_{lmn}^{i},\widehat{P}_{l^{\prime }m^{\prime }n^{\prime
}}^{i^{\prime }}\right] =i\hbar \delta ^{ii^{\prime }}\delta _{ll^{\prime
}}\delta _{mm^{\prime }}\delta _{nn^{\prime }},\quad \left[ \widehat{U}%
_{lmn}^{i},\widehat{U}_{l^{\prime }m^{\prime }n^{\prime }}^{i^{\prime }}%
\right] =\left[ \widehat{P}_{lmn}^{i},\widehat{P}_{l^{\prime }m^{\prime
}n^{\prime }}^{i^{\prime }}\right] =0.  \label{29}
\end{equation}

The Hamiltonian \eqref{ham} becomes:
\textcolor{black}{
\begin{align}\label{Ham_D}
\widehat H^D_{lmn} & =\frac{1}{2M}(1+R^{2})\left[\left(\widehat{P}_{lmn}^{x}\right)^{2}+\left(\widehat{P}_{lmn}^{y}\right)^{2}+\left(\widehat{P}_{lmn}^{z}\right)^{2}\right]-\frac{R^{2}}{2M}\left(\widehat{P}_{lmn}^{z}\widehat{P}_{lmn}^{y}\right)-\frac{R^{2}}{2M}\left(\widehat{P}_{lmn}^{x}\widehat{P}_{lmn}^{z}\right)-\frac{R^{2}}{2M}\left(\widehat{P}_{lmn}^{y}\widehat{P}_{lmn}^{x}\right)\nonumber \\
 & +\frac{\omega_{0}R}{2}\biggl\{\left(\widehat{P}_{lmn}^{z}-\widehat{P}_{lmn}^{y}\right)\left[2\widehat{U}_{lmn}^{x}-\widehat{U}_{l-1mn}^{x}-\widehat{U}_{l+1mn}^{x}\right]+\left(\widehat{P}_{lmn}^{x}-\widehat{P}_{lmn}^{z}\right)\left[2\widehat{U}_{lmn}^{y}-\widehat{U}_{lm-1n}^{y}-\widehat{U}_{lm+1n}^{y}\right]\nonumber \\
 & +\left(\widehat{P}_{lmn}^{y}-\widehat{P}_{lmn}^{x}\right)\left[2\widehat{U}_{lmn}^{z}-\widehat{U}_{lmn-1}^{z}-\widehat{U}_{lmn+1}^{z}\right]\biggr\}\nonumber \\
 & +\frac{M\omega_{0}^{2}}{2}\biggl\{\left(\widehat{U}_{lmn}^{x}-\widehat{U}_{l-1mn}^{x}\right)^{2}+\left(\widehat{U}_{l+1mn}^{x}-\widehat{U}_{lmn}^{x}\right)^{2}+\left(\widehat{U}_{lmn}^{y}-\widehat{U}_{lm-1n}^{y}\right)^{2}+\left(\widehat{U}_{lm+1n}^{y}-\widehat{U}_{lmn}^{y}\right)^{2}\nonumber \\
 & +\left(\widehat{U}_{lmn1}^{z}-\widehat{U}_{lmn-1}^{z}\right)^{2}+\left(\widehat{U}_{lmn+1}^{z}-\widehat{U}_{lmn1}^{z}\right)^{2}\biggr\},
\end{align}} where we have defined \textcolor{black}{$\omega_{\theta}=
\frac{\hbar}{M \theta}$} and $R=\frac{\omega_{0}}{\omega_{\theta}}$.

The equations of motion are obtained via Heisenberg's equations
\begin{equation}
i\hbar \frac{d\widehat{O}\left( t\right) }{dt}=\left[ \widehat{O}\left(
t\right) ,\widehat H\right] ,  \label{28}
\end{equation}%
where the generic operator stands for either the displacements or momenta of
the atom $(lmn)$, and $\widehat H$ is $\widehat H^D_{lmn}$ given by %
\eqref{Ham_D}.

The equations of motion for disordered atoms are
\begin{eqnarray}  \label{eom_D}
\ddot{\widehat{U}}_{lmn}^{x}=-\omega _{0}^{2}\left( 2\widehat{U}_{lmn}^{x}-%
\widehat{U}_{l-1mn}^{x}-\widehat{U}_{l+1mn}^{x}\right) +\frac{\omega _{0}R}{2%
}\left[ \left( 4\dot{\widehat{U}}_{lmn}^{y}-\dot{\widehat{U}}_{lm-1n}^{y}-%
\dot{\widehat{U}}_{lm+1n}^{y}\right) -\left( 4\dot{\widehat{U}}_{lmn}^{z}-%
\dot{\widehat{U}}_{lmn-1}^{z}-\dot{\widehat{U}}_{lmn+1}^{z}\right) \right],%
\cr \ddot{\widehat{U}}_{lmn}^{y}=-\omega _{0}^{2}\left( 2\widehat{U}%
_{lmn}^{y}-\widehat{U}_{lm-1n}^{y}-\widehat{U}_{lm+1n}^{y}\right) +\frac{%
\omega _{0}R}{2}\left[ \left( 4\dot{\widehat{U}}_{lmn}^{z}-\dot{\widehat{U}}%
_{lmn-1}^{z}-\dot{\widehat{U}}_{lmn+1}^{z}\right) -\left( 4\dot{\widehat{U}}%
_{lmn}^{x}-\dot{\widehat{U}}_{l-1mn}^{x}-\dot{\widehat{U}}%
_{l+1mn}^{x}\right) \right],\cr\ddot{\widehat{U}}_{lmn}^{z}=-\omega
_{0}^{2}\left( 2\widehat{U}_{lmn}^{z}-\widehat{U}_{lmn-1}^{z}-\widehat{U}%
_{lmn+1}^{z}\right) +\frac{\omega _{0}R}{2}\left[ \left( 4\dot{\widehat{U}}%
_{lmn}^{y}-\dot{\widehat{U}}_{lm-1n}^{y}-\dot{\widehat{U}}%
_{lm+1n}^{y}\right) -\left( 4\dot{\widehat{U}}_{lmn}^{z}-\dot{\widehat{U}}%
_{lmn-1}^{z}-\dot{\widehat{U}}_{lmn+1}^{z}\right) \right] ,
\end{eqnarray}
where we kept the terms up to the second order in $R$.

Any function in a space formed by a periodic arrangement of atoms must
satisfy periodic boundary conditions, the Born--von K\'{a}rm\'{a}n boundary
conditions. It is important to highlight that upon the shift of coordinates
applied to the quantum Hamiltonian, the disorder effects emerge as a new
interaction terms added to the ordinary Hamiltonian of the simple cubic
lattice. Therefore, the periodicity of the lattice is maintained, and the
Born--von K\'{a}rm\'{a}n boundary conditions can be freely applied.

Since the reciprocal lattice of a simple cubic lattice is another simple
cubic lattice, one can expand in Fourier series the operators%
\begin{equation}
\hat U^i_{lmn}=\sum_{\mathbf{k}}\hat{\mathcal{U}}^{i}_{\mathbf{k}}\exp \left[
i\left( \omega_{\mathbf{k}} t+a(lk_x+mk_y+n k_z) \right)\right] ,  \label{30}
\end{equation}%
in the above expression, $k_{i}$ represent the wave vectors and $\omega_{
\mathbf{k}}$ the vibrational frequencies.

Replacing the Ansatz \eqref{30} into the equations of motion \eqref{eom_D},
we find the saecular equation for the disordered atoms:
\begin{equation}
\left\vert
\begin{array}{ccc}
\omega_{\mathbf{k}} ^{2}-4\omega _{0}^{2}\sin ^{2}\left( \frac{ak_{x}}{2}%
\right) & i\omega_{\mathbf{k}} \omega _{0}R\left[ 1+2\sin ^{2}\left( \frac{%
ak_{y}}{2}\right) \right] & -i\omega_{\mathbf{k}} \omega _{0}R\left[ 1+2\sin
^{2}\left( \frac{ak_{z}}{2}\right) \right] \\
-i\omega_{\mathbf{k}} \omega _{0}R\left[ 1+2\sin ^{2}\left( \frac{ak_{x}}{2}%
\right) \right] & \omega_{\mathbf{k}} ^{2}-4\omega _{0}^{2}\sin ^{2}\left(
\frac{ak_{y}}{2}\right) & i\omega_{\mathbf{k}} \omega _{0}R\left[ 1+2\sin
^{2}\left( \frac{ak_{z}}{2}\right) \right] \\
i\omega_{\mathbf{k}} \omega _{0}R\left( 1+2\sin ^{2}\left( \frac{ak_{x}}{2}%
\right) \right) & -i\omega_{\mathbf{k}} \omega _{0}R\left[ 1+2\sin
^{2}\left( \frac{ak_{y}}{2}\right) \right] & \omega_{\mathbf{k}}
^{2}-4\omega _{0}^{2}\sin ^{2}\left( \frac{ak_{z}}{2}\right)%
\end{array}%
\right\vert =0 .  \label{saec_eq_disordered}
\end{equation}%
As explained Sect. III, the only physically relevant directions for our
model are the directions of coordinate axes. The isotropization of the model is
performed by taking the dispersion relations in the direction $Oz$ and
replicating it by rotational symmetry to all the directions of the Cartesian
system, namely by the replacement $k_z\to |{\mathbf{k}}|$.

\textcolor{black}{Upon isotropization, the solutions of equation \eqref{saec_eq_disordered}
(for $k_x=k_y\equiv 0$ and $k_z\to {\mathbf{k}}$), are:
\begin{eqnarray}  \label{omega_D}
\omega _{{\mathbf{k}}\pm }^{\mathrm{D}} &=&\frac{\omega _{0}}{2}\left[
6R^{2}+8\left( 1+R^{2}\right) \sin ^{2}\left( \frac{a|{\mathbf{k}}|}{2}%
\right) \pm 2\left[ -7R^{4}-32R^{2}-16\right. \right.  \label{32} \\
&&\left. \left. +\left( 56R^{4}+72R^{2}+32\right) \sin ^{2}\left( \frac{a|{%
\mathbf{k}}|}{2}\right) +\left( 16R^{4}+32R^{2}+16\right) \left( 1-2\sin
^{2}\left( \frac{a|{\mathbf{k}}|}{2}\right) +\sin ^{4}\left( \frac{a|{%
\mathbf{k}}|}{2}\right) \right) \right] ^{\frac{1}{2}}\right] ^{\frac{1}{2}},
\notag
\end{eqnarray}
where the minus (plus) sign is related to the degenerated acoustic (optical)
branch.  In the small angle approximation (i.e. $\sin (ak_{i})\approx ak_{i}$%
), this expressions becomes
\begin{equation}
\omega _{\mathbf{k}\pm}^{\mathrm{D-approx}}=\frac{\omega _{0}}{2}\left[
6R^{2}+2\left( 1+R^{2}\right) a^{2}|{\mathbf{k}}|^{2}\pm 2\sqrt{%
9R^{4}+2R^{2}\left( 3R^{2}+1\right) a^{2}|{\mathbf{k}}|^{2}+\left(
R^{2}+1\right) ^{2}a^{4}|{\mathbf{k}}|^{4}}\right]   \label{omega_D_app}
\end{equation}%
preserving the relation of the minus (plus) sign for the acoustic (optical)
branch.}

\begin{figure}[h]
\subfloat[Subfigure 1 list of figures text][Dispersion relations for disordered atoms]  {\
\includegraphics[width=6.5cm]{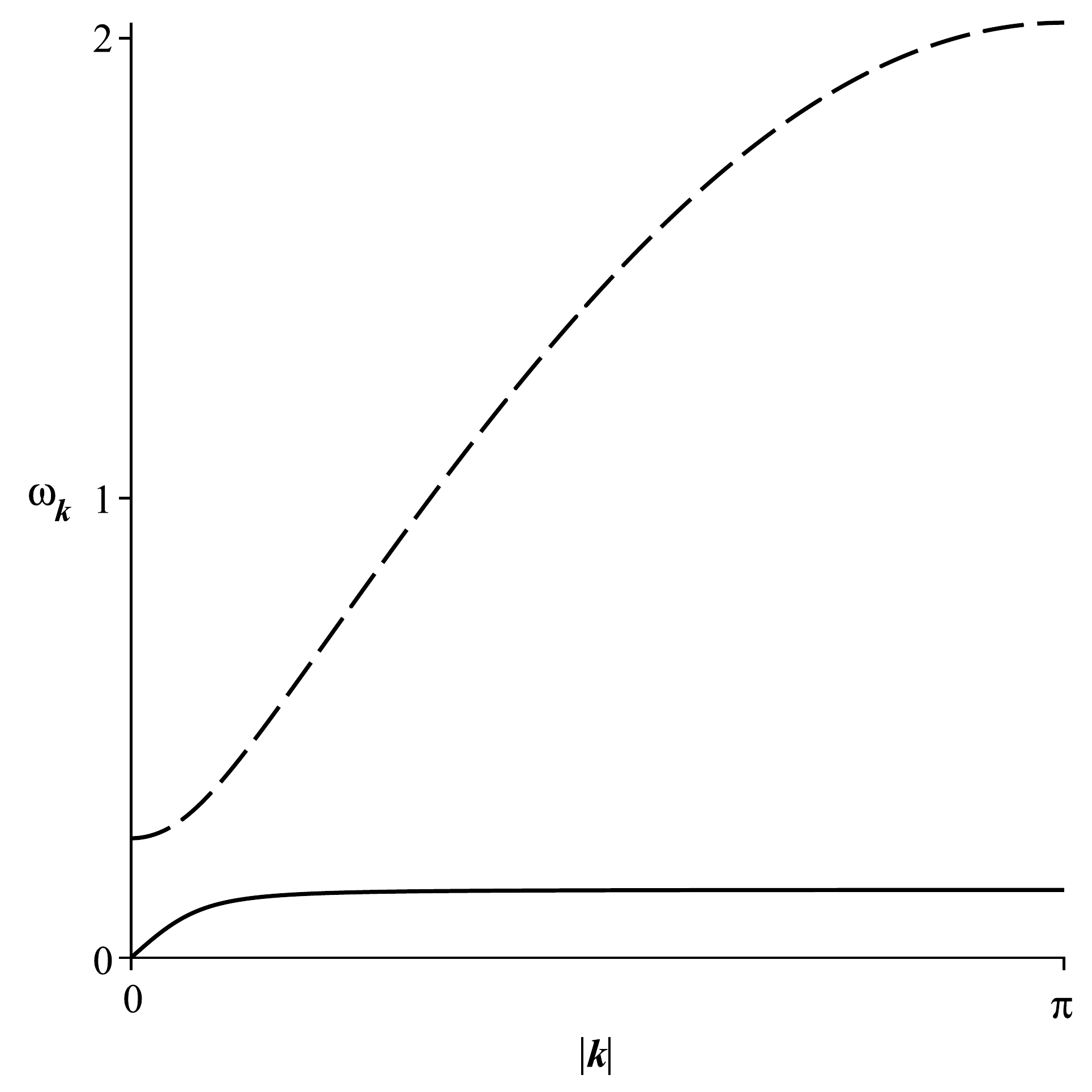}} ~
\subfloat[Subfigure 1 list of figures text][Dispersion relations for ordered atoms]  {\
\includegraphics[width=6.5cm]{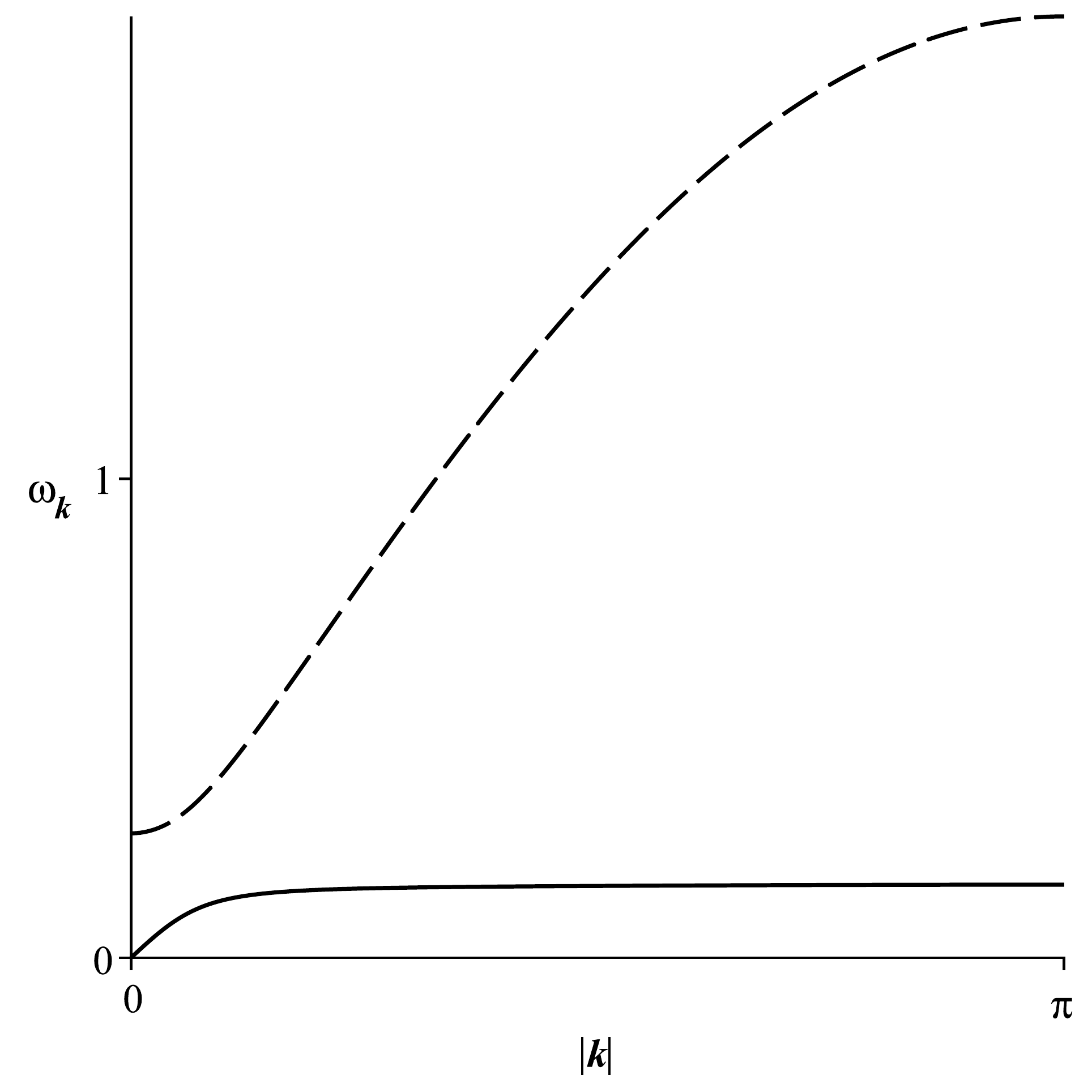}} \newline
\subfloat[Subfigure 1 list of figures text][Zoom at the acoustic branch]  {\
\includegraphics[width=6.5cm]{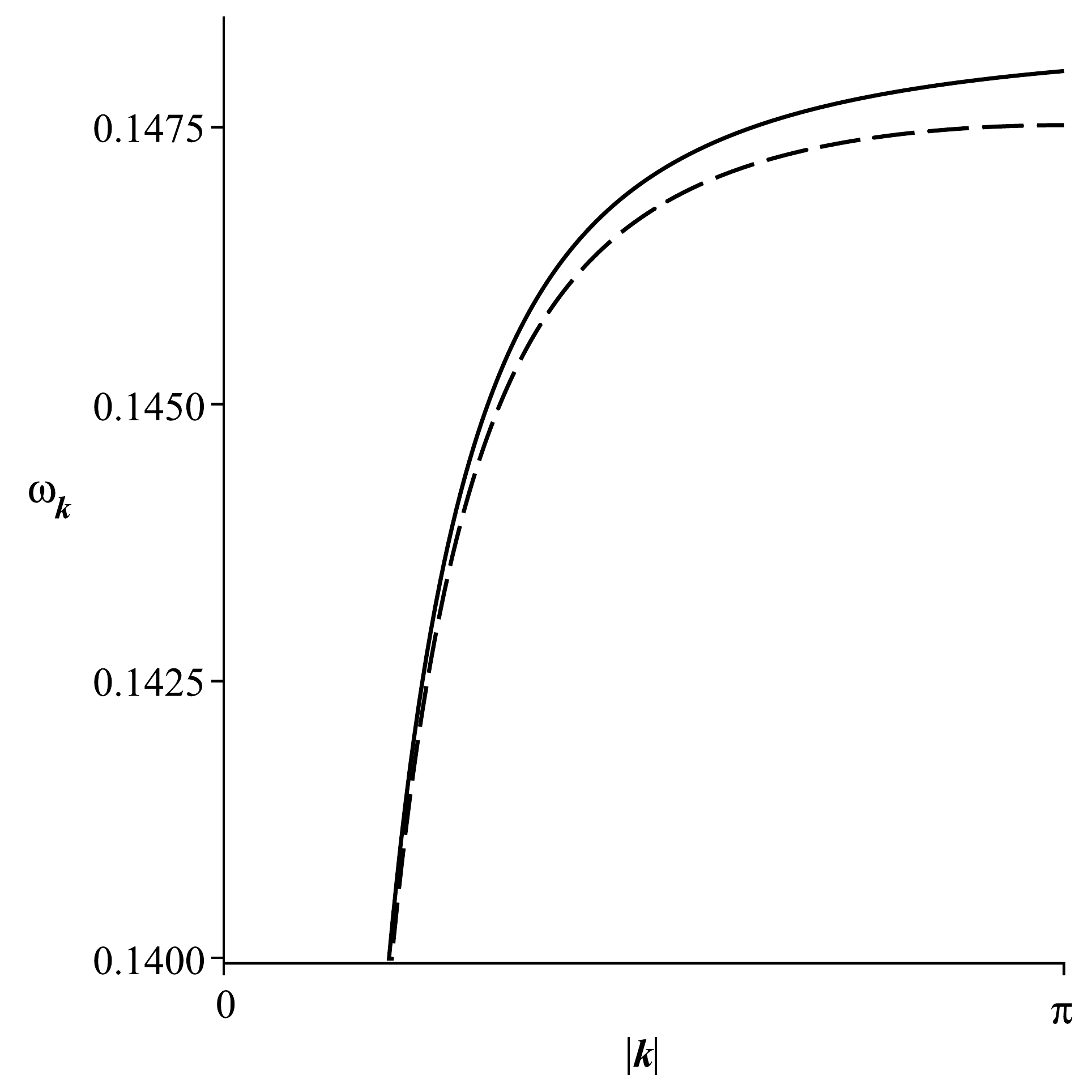}} ~
\subfloat[Subfigure 1 list of figures text][Zoom at the acoustic branch]  {\
\includegraphics[width=6.5cm]{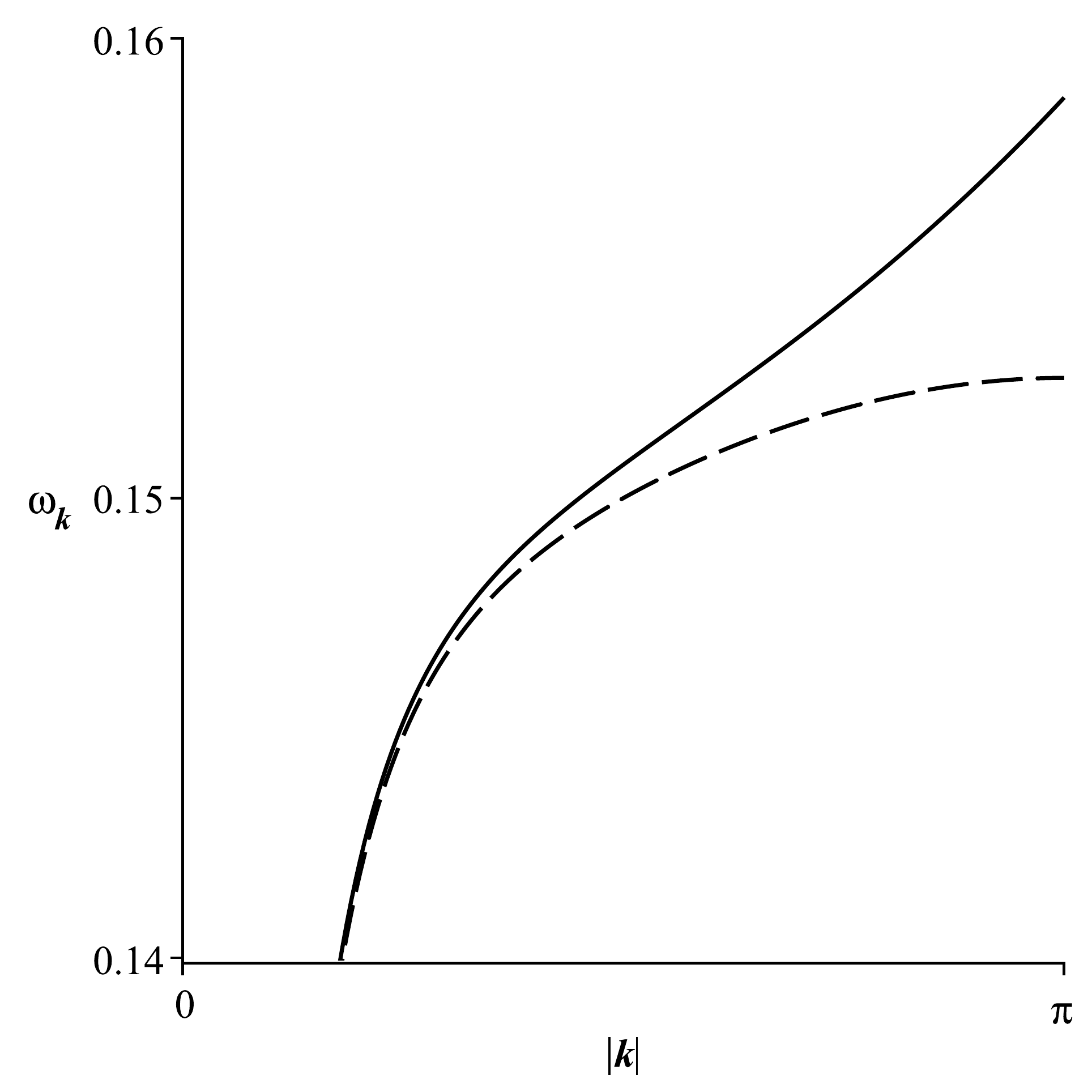}}
\caption{\textbf{Dispersion relations for $R = 0.15$.}
Panels (a) and (b) present the dispersion relations for disordered and
ordered atoms, respectively. The degenerated acoustic branches are represented by the solid lines, while the optical modes are represented by
dashed lines. Panels (c) and (d) present a comparison between the
approximated (solid line) and exact (dashed line) solutions for the acoustic branch (a zoom in the end
of the Brillouin zone, in order to reveal their slight discrepancy).} \label{completapprox}
\end{figure}

Both functions \eqref{omega_D} and \eqref{omega_D_app} are monotonically
increasing within the Brillouin zone, with the maximum at $ak_z=\pi$. In
Fig.~\ref{completapprox} are depicted the dispersion relations for $R=0.15$.
We note the characteristic plateau in $\omega _{\mathbf{k}}$, which leads to
the broadened van Hove singularity in the VDOS. The small angle
approximation gives a very good estimate of the full solution for small $R$,
therefore we work further with the approximate dispersion relation %
\eqref{omega_D_app}, which is the solution of the equation
\begin{equation}
\omega_{\mathbf{k}} ^{4}+\omega _{0}^{2}\left[ \omega _{0}^{2}R^{2}-\omega_{%
\mathbf{k}} ^{2}\left( 1+R^{2}\right) \right] a^{2}|\mathbf{k}|^2 -3\omega_{%
\mathbf{k}} ^{2}\omega _{0}^{2}R^{2}=0.  \label{disordered_se}
\end{equation}
We may put equation \eqref{disordered_se} in the form
\begin{equation}
k_{x}^{2}+k_{y}^{2}+k_{z}^{2}=\frac{1}{a^{2}}\frac{3\omega_{\mathbf{k}}
^{2}R^{2}-\frac{\omega_{\mathbf{k}} ^{4}}{\omega _{0}^{2}}}{\omega
_{0}^{2}R^{2}-\omega_{\mathbf{k}} ^{2}\left( 1+R^{2}\right) }
\label{volume_disordered}
\end{equation}%
and identify the isofrequency surfaces $\omega_{\mathbf{k}}=\omega=\text {%
const.}$ in the ${\mathbf{k}}$-space as spheres of radius $\frac{1}{a^{2}}%
\frac{3\omega ^{2}R^{2}-\frac{\omega ^{4}}{\omega _{0}^{2}}}{\omega
_{0}^{2}R^{2}-\omega ^{2}\left( 1+R^{2}\right) }$. The density of states is
obtained by taking the derivative of the volume of the sphere with respect
to $\omega$ and subsequently dividing by the volume of one cell in the ${%
\mathbf{k}}$-space, $(2\pi)^3/V$, where $V$ is the volume of the system in
the direct space. The result is
\begin{equation}
g_{\mathrm{D}}\left( \omega \right) =\frac{3}{2} \frac{V}{ \left( 2\pi
\right) ^{3}}\frac{4\pi \omega ^{2}}{\omega _{0}^{3}a^{3}}\sqrt{\left\vert
\frac{3-\frac{\omega ^{2}}{\omega _{0}^{2}R^{2}}}{1-\frac{\omega ^{2}}{%
\omega _{0}^{2}R^{2}}}\right\vert }\left( \frac{\omega ^{2}}{\omega
_{0}^{2}R^{2}}\right) \left\vert \frac{\frac{\omega ^{2}}{\omega
_{0}^{2}R^{2}}-2-\frac{3}{2}\frac{\omega ^{2}}{\omega _{0}^{2}}}{\left( 1-%
\frac{\omega ^{2}}{\omega _{0}^{2}R^{2}}\right) ^{2}}\right\vert ,
\label{38}
\end{equation}%
where the prefactor 3/2 accounts for the three acoustic branches and the
fact that only half of the atoms are disordered.

\section{Equations of motion and dispersion relations for ordered atoms}

In a similar manner as described in the preceding section, we proceed
with the ordered atoms. If the atom $(lmn)$ is a ordered one, we have
\begin{eqnarray}
\widehat{u}^i_{lmn}&=&\widehat U^i_{lmn}, \ \ \ i=x,y,z,
\end{eqnarray}
while its nearest neighbors are disordered and the shift of noncommutative
coordinates has to be applied to them, as follows:
\begin{eqnarray}
\widehat{u}^x_{l\pm1mn}=\widehat U^x_{l\pm1mn}-\frac{1}{2\hbar}%
\theta_{xj}\widehat P_{l\pm1mn}^{j}, \ \ j=y,z,\cr \widehat{u}%
^y_{lm\pm1n}=\widehat U^y_{lm\pm1n}-\frac{1}{2\hbar}\theta_{yj}\widehat
P_{lm\pm1n}^{j}, \ \ j=z,x,\cr \widehat{u}^z_{lmn\pm1}=\widehat
U^z_{lmn\pm1}-\frac{1}{2\hbar}\theta_{zj}\widehat P_{lmn\pm1}^{j}, \ \
j=x,y.
\end{eqnarray}
As for the momenta, $\widehat{p}_{lmn}^{i}=\widehat P_{lmn}^{i} . $ The
Hamiltonian \eqref{ham} becomes:
\textcolor{black}{\begin{align}\label{Ham_O}
\widehat H^O_{lmn} & =\frac{1}{2M}\left[\left(\widehat{P}_{lmn}^{x}\right)^{2}+\left(\widehat{P}_{lmn}^{y}\right)^{2}+\left(\widehat{P}_{lmn}^{z}\right)^{2}\right] \\
 & +\frac{M\omega_{0}^{2}}{2}\left[\left(\widehat{U}_{lmn}^{x}-\widehat{U}_{l-1mn}^{x}-\frac{\theta}{2}\left(\widehat{P}_{l-1mn}^{z}-\widehat{P}_{l-1mn}^{y}\right)\right)^{2}+\left(\widehat{U}_{l+1mn}^{x}+\frac{\theta}{2}\left(\widehat{P}_{l+1mn}^{z}-\widehat{P}_{l+1mn}^{y}\right)-\widehat{U}_{lmn}^{x}\right)^{2}\right]\nonumber \\
 & +\frac{M\omega_{0}^{2}}{2}\left[\left(\widehat{U}_{lmn}^{y}-\widehat{U}_{lm-1n}^{y}-\frac{\theta}{2}\left(\widehat{P}_{lm-1n}^{x}-\widehat{P}_{lm-1n}^{z}\right)\right)^{2}+\left(\widehat{U}_{lm+1n}^{y}+\frac{\theta}{2}\left(\widehat{P}_{lm+1n}^{x}-\widehat{P}_{lm+1n}^{z}\right)-\widehat{U}_{lmn}^{y}\right)^{2}\right]\nonumber \\
 & +\frac{M\omega_{0}^{2}}{2}\left[\left(\widehat{U}_{lmn}^{z}-\widehat{U}_{lmn-1}^{z}-\frac{\theta}{2}\left(\widehat{P}_{lmn-1}^{y}-\widehat{P}_{lmn-1}^{x}\right)\right)^{2}+\left(\widehat{U}_{lmn+1}^{z}+\frac{\theta}{2}\left(\widehat{P}_{lmn+1}^{y}-\widehat{P}_{lmn+1}^{x}\right)-\widehat{U}_{lmn}^{z}\right)^{2}\right].\nonumber
\end{align}}

Proceeding as in the case of disordered atoms, we find the equations of
motion up to the second order in $R$:
\begin{eqnarray}  \label{O_solution}
\ddot{\widehat U}^x_{lmn} & =&-\omega_{0}^{2}\left(2{\widehat U}^x_{lmn}-{%
\widehat U}^x_{l-1mn}-{\widehat U}^x_{l+1mn}\right)+\frac{\omega_{0}R}{2}%
\biggl\{\dot{\widehat U}^z_{l-1mn}+\dot{\widehat U}^z_{l+1mn}-\dot{\widehat U%
}^y_{l-1mn}-\dot{\widehat U}^y_{l+1mn}\biggr\}\cr & +&\frac{%
\omega_{0}^{2}R^{2}}{4}\biggl\{-2\left[2{\widehat U}^x_{l-1mn}-{\widehat U}%
^x_{l-2mn}-{\widehat U}^x_{lmn}\right]-2\left[2{\widehat U}^x_{l+1mn}-{%
\widehat U}^x_{l+2mn}-{\widehat U}^x_{lmn}\right]\cr & +&\left[2{\widehat U}%
^y_{l-1mn}-{\widehat U}^y_{l-1m-1n}-{\widehat U}^y_{l-1m+1n}\right]+\left[2{%
\widehat U}^y_{l+1mn}-{\widehat U}^y_{l+1m-1n}-{\widehat U}^y_{l+1m+1n}%
\right]\cr & +&\left[2{\widehat U}^z_{l-1mn}-{\widehat U}^z_{l-1mn-1}-{%
\widehat U}^z_{l-1mn+1}\right]+\left[2{\widehat U}^z_{l+1mn}-{\widehat U}%
^z_{l+1m+2n-1}-{\widehat U}^z_{l+1mn+1}\right]\biggr\}.
\end{eqnarray}
The saecular equation for the ordered atoms is given by
\begin{equation}
\det|A_{ij}|=0,
\end{equation}
where the elements of the matrix $A_{ij}$ are%
\begin{eqnarray*}
&&A_{11}=\omega_{\mathbf{k}} ^{2}-4\omega _{0}^{2}\sin ^{2}\left( \frac{%
ak_{x}}{2}\right) +2\omega _{0}^{2}R^{2}\left[ 2\sin ^{2}\left( \frac{ak_{x}%
}{2}\right) -\sin ^{2}\left( ak_{x}\right) \right], \\
&&A_{12}=-i\omega _{0}\omega_{\mathbf{k}} R\left[ 1-2\sin ^{2}\left( \frac{%
ak_{x}}{2}\right) \right] +2\omega _{0}^{2}R^{2}\left[ \sin ^{2}\left( \frac{%
ak_{y}}{2}\right) -2\sin ^{2}\left( \frac{ak_{x}}{2}\right) \sin ^{2}\left(
\frac{ak_{y}}{2}\right) \right], \\
&&A_{13}=i\omega _{0}\omega_{\mathbf{k}} R\left[ 1-2\sin ^{2}\left( \frac{%
ak_{x}}{2}\right) \right] +2\omega _{0}^{2}R^{2}\left[ \sin ^{2}\left( \frac{%
ak_{z}}{2}\right) -2\sin ^{2}\left( \frac{ak_{x}}{2}\right) \sin ^{2}\left(
\frac{ak_{z}}{2}\right) \right],
\end{eqnarray*}%
\begin{eqnarray*}
A_{21} &=&i\omega_{\mathbf{k}} \omega _{0}R\left( 1-2\sin ^{2}\left( \frac{%
ak_{y}}{2}\right) \right) +2\omega _{0}^{2}R^{2}\left[ \sin ^{2}\left( \frac{%
ak_{x}}{2}\right) -2\sin ^{2}\left( \frac{ak_{x}}{2}\right) \sin ^{2}\left(
\frac{ak_{y}}{2}\right) \right], \\
A_{22} &=&\omega_{\mathbf{k}} ^{2}-4\omega _{0}^{2}\sin ^{2}\left( \frac{%
ak_{y}}{2}\right) +2\omega _{0}^{2}R^{2}\left[ 2\sin ^{2}\left( \frac{ak_{y}%
}{2}\right) -\sin ^{2}\left( ak_{y}\right) \right], \\
A_{23} &=&-i\omega_{\mathbf{k}} \omega _{0}R\left[ 1-2\sin ^{2}\left( \frac{%
ak_{y}}{2}\right) \right] +2\omega _{0}^{2}R^{2}\left[ \sin ^{2}\left( \frac{%
ak_{z}}{2}\right) -2\sin ^{2}\left( \frac{ak_{y}}{2}\right) \sin ^{2}\left(
\frac{ak_{z}}{2}\right) \right],
\end{eqnarray*}%
\begin{eqnarray}
A_{31} &=&-i\omega_{\mathbf{k}} \omega _{0}R\left[ 1-2\sin ^{2}\left( \frac{%
ak_{z}}{2}\right) \right] +2\omega _{0}^{2}R^{2}\left[ \sin ^{2}\left( \frac{%
ak_{x}}{2}\right) -2\sin ^{2}\left( \frac{ak_{x}}{2}\right) \sin ^{2}\left(
\frac{ak_{z}}{2}\right) \right],  \label{saec_eq_ordered} \\
A_{32} &=&i\omega_{\mathbf{k}} \omega _{0}R\left[ 1-2\sin ^{2}\left( \frac{%
ak_{z}}{2}\right) \right] +2\omega _{0}^{2}R^{2}\left[ \sin ^{2}\left( \frac{%
ak_{y}}{2}\right) -2\sin ^{2}\left( \frac{ak_{y}}{2}\right) \sin ^{2}\left(
\frac{ak_{z}}{2}\right) \right],  \notag \\
A_{33} &=&\omega_{\mathbf{k}} ^{2}-4\omega _{0}^{2}\sin ^{2}\left( \frac{%
ak_{z}}{2}\right) +2\omega _{0}^{2}R^{2}\left[ 2\sin ^{2}\left( \frac{ak_{z}%
}{2}\right) -\sin ^{2}\left( ak_{z}\right) \right].  \notag
\end{eqnarray}
Upon isotropization and in the small angle approximation, the saecular
equation for ordered atoms becomes
\begin{equation}
\omega_{\mathbf{k}} ^{4}-\omega _{0}^{2}a^{2}\left( \omega
_{0}^{2}R^{2}-\omega_{\mathbf{k}} ^{2}\right) |\mathbf{k}|^2 +3\omega_{%
\mathbf{k}} ^{2}\omega _{0}^{2}R^{2}=0,  \label{ordered_se}
\end{equation}
or
\begin{equation}
k_{x}^{2}+k_{y}^{2}+k_{z}^{2}=\frac{1}{a^{2}}\frac{3\omega ^{2}R^{2}-\frac{%
\omega ^{4}}{\omega _{0}^{2}}}{\omega _{0}^{2}R^{2}-\omega ^{2}}.
\label{volume_ordered}
\end{equation}
\textcolor{black}{The solutions for the saecular equation coming from (\ref{saec_eq_ordered}) are:
\begin{eqnarray}  \label{omega_O}
\omega _{{\mathbf{k}}\pm }^{\mathrm{O}} &=&\frac{\omega _{0}}{2}\left[
6R^{2}+8\sin ^{2}\left( \frac{a|{\mathbf{k}}|}{2}\right) \left[ 1-2R^{2}\sin
^{2}\left( \frac{a|{\mathbf{k}}|}{2}\right) \right] \right.   \label{34} \\
&&\left. \pm 2\left[ 64R^{2}\sin ^{6}\left( \frac{a|{\mathbf{k}}|}{2}\right) %
\left[ R^{2}\sin ^{2}\left( \frac{a|{\mathbf{k}}|}{2}\right) -1\right]
+\left( 16-48R^{4}\right) \sin ^{4}\left( \frac{a|{\mathbf{k}}|}{2}\right)
+8R^{2}\sin ^{2}\left( \frac{a|{\mathbf{k}}|}{2}\right) +9R^{4}\right] ^{%
\frac{1}{2}}\right] ^{\frac{1}{2}}.  \notag
\end{eqnarray}%
Accordingly, in the small angle approximation, this expression is rewritten as:
\begin{eqnarray}
\omega _{{\mathbf{k}}\pm }^{\mathrm{O-approx}} &=&\frac{\omega _{0}}{2}\left[
6R^{2}+2a^{2}|{\mathbf{k}}|^{2}\left( 1-\frac{R^{2}}{2}a^{2}|{\mathbf{k}}%
|^{2}\right) \right.   \label{omega_O_app}\\
&&\left. \pm 2\sqrt{R^{2}a^{6}|{\mathbf{k}}|^{6}\left( \frac{R^{2}}{4}a^{2}|{%
\mathbf{k}}|^{2}-1\right) +\left( 1-3R^{4}\right) a^{4}|{\mathbf{k}}%
|^{4}+2R^{2}a^{2}|{\mathbf{k}}|^{2}+9R^{4}}\right] ^{\frac{1}{2}}. \notag
\end{eqnarray}}
Similarly to the case of the disordered atoms, the ordered atoms contribution to the density of states is found to be:
\begin{equation}
g_{O}\left( \omega \right) =\textcolor{black}3\frac{V}{(2\pi)^2
}{\frac{1}{a^3}}\frac{\omega ^{2}}{\omega _{0}^{3}}\sqrt{\left\vert \frac{3-\frac{\omega
^{2}}{\omega _{0}^{2}R^{2}}}{1-\frac{\omega ^{2}}{\omega _{0}^{2}R^{2}}}%
\right\vert }. \label{g_order}
\end{equation}

\section{Specific heat of the glass}

\textcolor{black}{The complete glass DOS in the small angle approximation is
equal to the sum of the disordered and ordered atoms expressions for DOS, namely }
\textcolor{black} {%
\begin{equation}  \label{g_glass}
g_{glass}\left( \omega \right) =g_{O}\left( \omega \right) \left[ 1+\frac{%
\omega ^{2}}{\omega _{0}^{2}R^{2}}\left\vert \frac{\frac{\omega ^{2}}{\omega
_{0}^{2}R^{2}}-2-\frac{3}{2}\frac{\omega ^{2}}{\omega _{0}^{2}}}{\left( 1-%
\frac{\omega ^{2}}{\omega _{0}^{2}R^{2}}\right) ^{2}}\right\vert \right],
\end{equation}%
where $g_{O}\left( \omega \right)$ is the contribution to VDOS from the
ordered atoms given by \ref{g_order}.}

The reduced specific heat is determined from the following expression

\textcolor{black}{
\begin{equation}\label{specific_heat_app}
\frac{C}{T^3}=\int_0^{\omega_{\rm{max}}} \frac{ \hbar^2 \omega^2}{k_{B}T^5}\frac{N_A}{Z} \frac{e^{\hbar\omega/k_{B}T}}{(e^{\hbar\omega/k_{B}T}-1)^2} g_{glass}(\omega) d\omega,
\end{equation}}  in which $Z$ is the number of formula units per unit cell
(in our model, $Z=1$) and $k_B$ is the Boltzmann constant. We take $\omega _{%
\mathrm{max}}=\sqrt 3\omega _{0}R$ as the frequency of the optical branches
at $\mathbf{k}=0$. We adopt this value for the maximum frequency because our small angle approximation makes the acoustic branch frequency at the end of the Brillouin zone slightly larger than the frequency corresponding to the exact solution. In this way we insure that we integrate over the whole acoustic branch.

In order to establish an order of magnitude for the noncommutativity
parameter $\theta $, implicitly a measure of nonlocality, one has to resort
to the available experimental data: from the match between the theoretical
prediction with the experimental data for the reduced specific heat one can
assign that the unique pair of values of $R$ and $\omega _{0}$ used for the
fit gives the value for the frequency of divergence $\omega _{\mathrm{div}}$. From this set of parameters one can establish the value for $\theta $
by means of $\omega _{\theta }=\frac{\hbar }{m\theta }$.


\begin{thebibliography}{99}


\bibitem{berthier} L. Berthier and G. Biroli, ``Theoretical
perspective on the glass transition and amorphous materials.'' Rev.
Mod. Phys. {\bf 83}, 587  (2011).

\bibitem{livro} J. M. Ziman, ``Models of disorder: the theoretical physics
of homogeneously disordered systems'' (CUP Archive, 1979).\\
S. R. Elliott, ``Physics of Amorphous Materials'' (Longman, New
York, 1990).
% Book  phillips

\bibitem{shintani} H. Shintani and H. Tanaka, ``Universal link
between the boson peak and transverse phonons in glass'', Nature materials
{\bf7}, 870 (2008).

\bibitem{chumakov} A. I. Chumakov, et al., \textquotedblleft Equivalence of
the boson peak in glasses to the transverse acoustic van Hove singularity in
crystals.\textquotedblright\ Phys. Rev. Lett. {\bf106}, 225501 (2011).

\bibitem{Brink} T. Brink, L. Koch, and K. Albe, ``Structural origins of the boson peak in metals:
From high-entropy alloys to metallic glasses'', Phys. Rev. B {\bf94}, 224203 (2016).

\bibitem{Karpov}V. G. Karpov, M. I. Klinger, and F. N. Ignatiev, Sov. Phys. JETP {\bf57}, 439 (1983).

\bibitem{Laird}B. B. Laird and H. R. Schober, ``Localized Low-Frequency Vibrational Modes in a Simple Model Glass'', Phys. Rev. Lett. {\bf66}, 636 (1991).

\bibitem{27}W. Schirmacher, G. Diezemann and C. Ganter,  Harmonic vibrational excitations in disordered solids
and the ‘boson peak’, Phys. Rev. Lett. {\bf81}, 136 (1998).
%\bibitem{28} W. Schirmacher, Thermal conductivity of glassy materials and the “boson peak”. Europhys. Lett. 73,
%892–898 (2006).
\bibitem{29} W. Schirmacher, G. Ruocco, and T. Scopigno, ``Acoustic attenuation in glasses and its relation with the
Boson peak'', Phys. Rev. Lett. {\bf98}, 025501 (2007).

\bibitem{21} S. Elliott, ``A unified model for the low-energy vibrational behaviour of amorphous solids'', Europhys.
Lett. {\bf19}, 201 (1992).


\bibitem{25} E. Duval, A. Boukenter, and T. Achibat,  ``Vibrational dynamics and the structure of glasses'', J. Phys.
Condens. Matter {\bf2}, 10227 (1990).


\bibitem{22} M. I. Klinger, ``Atomic quantum diffusion, tunnelling states and some related phenomena in
condensed systems'', Phys. Rep. {\bf94}, 184 (1983).

\bibitem{23} M. I. Klinger and A. M. Kosevich, ``Soft-mode dynamics model of boson peak and high frequency sound
in glasses: Inelastic Ioffe--Regel crossover and strong hybridization of excitations'', Phys. Lett. A {\bf295},
311 (2002).

\bibitem{24} U. Buchenau, Yu. M. Galperin, V. L. Gurevich, D. A. Parshin, M. A. Ramos, and H. R. Schober, ``Interaction of soft modes and sound waves in glasses'', Phys. Rev. B {\bf46},
2798 (1992).


\bibitem{34}H. Tanaka, ``Physical origin of the boson peak deduced from a two-order-parameter model of liquid'',
J. Phys. Soc. Jpn. {\bf70}, 1178 (2001).

\bibitem{30} T. Grigera, V. Martin-Mayor, G. Parisi, and P. Verrocchio, ``Phonon interpretation of the ‘boson peak’ in
supercooled liquids'', Nature {\bf422}, 289 (2003).


\bibitem{26}W. G\"otze and M. R. Mayr, ``Evolution of vibrational excitations in glassy systems'', Phys. Rev. E {\bf61},
587 (2000).
\bibitem{26'} M. Tokuyama, ``Statistical-mechanical theory of nonlinear density fluctuations near the glass transition'', Physica A {\bf395}, 31 (2014).


\bibitem{31} V. Lubchenko and P. G. Wolynes,  ``The origin of the boson peak and thermal conductivity plateau in
low-temperature glasses'', Proc. Natl Acad. Sci. USA {\bf100}, 1515 (2003).
%\bibitem{32} Wittmer, J. P., Tanguy, A., Barrat, J. L. and Lewis, L. Vibrations of amorphous, nanometric structures:
%When does continuum theory apply? Europhys. Lett. 57, 423–429 (2002).
\bibitem{33} L. Silbert, A.J. Liu, and S. Nagel, ``Vibrations and Diverging Length Scales Near the Unjamming Transition'', Phys. Rev. Lett. {\bf95}, 098301 (2005).

%M. Wyart, S.R. Nagel, and T. A. Witten, "Geometric origin of excess low-frequency vibrational modes
%in weakly connected amorphous solids", Europhys. Lett. 72, 486 (2005).

%\bibitem{Chem} ??? Chen, Ke, et al. "Low-frequency vibrations of soft colloidal
%glasses." Physical review letters 105.2 (2010): 025501.

%\bibitem{Pohl} ??? Pohl, Robert O., Xiao Liu, and EunJoo Thompson.
%"Low-temperature thermal conductivity and acoustic attenuation in amorphous
%solids." Reviews of Modern Physics 74.4 (2002): 991.

%\bibitem{kittel} Kittel, Charles. Introduction to solid state. Fouth
%Edition. John Wiley \& Sons, 1971.




\bibitem{Mechanics}
M. Chaichian, I. Merches, and A. Tureanu, ``Mechanics: An Intensive Course'' (Springer, Berlin Heidelberg, 2012).

%\bibitem{phillips} Phillips, James C. "Critical points and lattice vibration
%spectra." Physical Review 104.5 (1956): 1263.

\bibitem{Susskind-Bahcall}
S. Bahcall and L. Susskind, ``Fluid Dynamics, Chern-Simons Theory and the
Quantum Hall Effect'', Int. J. Mod. Phys. B {\bf5}, 2735 (1991).


\bibitem{Susskind} L.~Susskind, ``The Quantum Hall fluid and noncommutative
Chern-Simons theory,'' hep-th/0101029.


\bibitem{Jackiw1} R.~Jackiw, S.~Y.~Pi, and A.~P.~Polychronakos, \textquotedblleft
Noncommuting gauge fields as a Lagrange fluid,\textquotedblright\ Annals
Phys.\ \textbf{301} (2002) 157 [hep-th/0206014].

\bibitem{Jackiw2} R.~Jackiw, V.~P.~Nair, S.~Y.~Pi, and A.~P.~Polychronakos,
\textquotedblleft Perfect fluid theory and its
extensions,\textquotedblright\ J.\ Phys.\ A \textbf{37} (2004) R327
[hep-ph/0407101].

\bibitem{Polyn} A.~P.~Polychronakos, \textquotedblleft Non-commutative
Fluids,\textquotedblright\ Prog.\ Math.\ Phys.\ \textbf{53} (2007) 109
[arXiv:0706.1095 [hep-th]].

\bibitem{CDP_98}
M.~Chaichian, A.~Demichev, and P.~Pre\v{s}najder,
  ``Quantum field theory on noncommutative space-times and the persistence of ultraviolet divergences,''
  Nucl.\ Phys.\ B {\bf 567}, 360 (2000)
%  doi:10.1016/S0550-3213(99)00664-1
  [hep-th/9812180].

 \bibitem{Bigatti_Susskind}
D.~Bigatti and L.~Susskind,
  ``Magnetic fields, branes and noncommutative geometry,''
  Phys.\ Rev.\ D {\bf 62}, 066004 (2000)
%  doi:10.1103/PhysRevD.62.066004
  [hep-th/9908056].

\bibitem{Anca} M.~Chaichian, K.~Nishijima, and A.~Tureanu,
  ``An Interpretation of noncommutative field theory in terms of a quantum shift,''
  Phys.\ Lett.\ B {\bf 633}, 129 (2006)
%  doi:10.1016/j.physletb.2005.11.051
  [hep-th/0511094].

\bibitem{Tureanu} M.~Chaichian, M.~M.~Sheikh-Jabbari, and A.~Tureanu,
  ``Hydrogen atom spectrum and the Lamb shift in noncommutative QED,''
  Phys.\ Rev.\ Lett.\  {\bf 86}, 2716 (2001)
 % doi:10.1103/PhysRevLett.86.2716
  [hep-th/0010175].

\bibitem{Kantorovich}
L. Kantorovich, ``Quantum Theory of the Solid State: An Introduction'' (Springer, Berlin Heidelberg, 2004).




%\bibitem{Fiechtner} Fiechtner, Gregory J., and James R. Gord. "Absorption and the
%dimensionless overlap integral for two-photon excitation." Journal of
%Quantitative Spectroscopy and Radiative Transfer 68.5 (2001): 543-557.

\bibitem{queen} D.R. Queen, X. Liu., J. Karel, T.H. Metcalf, and F. Hellman,
``Excess Specific Heat in Evaporated Amorphous
Silicon'', Phys. Rev. Lett. \textbf{110}, 135901
(2013).


\bibitem{GeO2}
P. Richet, D. de Ligny, and E.F. Westrum Jr., ``Low-temperature heat capacity of GeO$_2$ and B$_2$O$_3$ glasses:
thermophysical and structural implications'', J. Non-Cryst. Solids {\bf 315}, 20 (2003).

\bibitem{BGS} K. Suekuni, M. A. Avila, K. Umeo, H. Fukuoka, S. Yamanaka, T. Nakagawa,  and  T. Takabatake, \textquotedblleft Simultaneous
structure and carrier tuning of dimorphic clathrate Ba$_{8}$Ga$_{16}$Sn$_{30}$,\textquotedblright\ Phys. Rev. B \textbf{77}, 235119  (2008).

%\bibitem{silica_data}K. Nakamura, Y. Takahashi, and T. Fujiwara, "Low-temperature excess heat capacity in fresnoite glass and %crystal", Scientific reports {\bf 4}, 6523 (2014).

%\bibitem{miguel}  T. P\'{e}rez-Casta\~{n}eda, R. J. Jim\'{e}nez-Riob\'{o}o,
%and M. A. Ramos,  \textquotedblleft Two-level systems and boson peak
%remain stable in 110-million-year-old amber glass\textquotedblright .
%Phys. Rev. Lett. 112, 165901 (2014) and its Supplemental Material.

\bibitem{work in progress}
T.R. Cardoso, R. Bufalo, and A. Tureanu, work in progress.


%\bibitem{Buh} Bufalo, R., T. R. Cardoso, and B. M. Pimentel. "K\"{a}ll\'{e}%
%n-Lehmann representation of noncommutative quantum electrodynamics."
%Physical Review D 89.8 (2014): 085010.

\end{thebibliography}
\end{document}